\documentclass{PoS}

\newcommand{\nn}{\nonumber}

\renewcommand{\(}{\left(}
\renewcommand{\)}{\right)}

\title{TMDs in Laguerre polynomial basis}

\ShortTitle{TMDs in Laguerre polynomial basis}

\author{\speaker{Alexey Vladimirov}
\\
         Department of Astronomy and Theoretical Physics, Lund University,\\  S\"olvegatan 14A, S 223 62 Lund, Sweden\\
        E-mail: \email{vladimirov.aleksey@gmail.com}}

\abstract{We suggest the modification of the standard approach to TMDs. The modification consists in the consideration of the small $b_T$
operator product expansion in the different operator basis. Instead of power expansion we suggest to use the Laguerre polynomial expansion.
Within such a scheme the first term of OPE saturates TMDs in the wider range of $b_T$ in comparison to the power expansion that decreases the
significance of non-perturbative factor at small $b_T$. The presented modification does not violate any TMD properties and can be used within
any formulation of TMD factorization.}

\FullConference{XXII. International Workshop on Deep-Inelastic Scattering and Related Subjects,\\
        28 April - 2 May 2014\\
        Warsaw, Poland}

\begin{document}

\section{TMDs with maximum perturbative content}

Transverse momentum dependent (TMD) parton distribution functions (PDFs) and fragmentation functions (FFs) (we will refer them collectively as
TMDs) are universal functions which accumulate information about intrinsic structure of hadrons. TMDs express the leading behavior of processes
with two detected states in the range of intermediate transverse momentum $Q\gg q_{hT}\gg\Lambda_{QCD}$. The examples of such processes are
Drell-Yan process, semi-inclusive deep inelastic scattering (SIDIS), and $e^+e^-$-annihilation to two jets. The typical expression for the
hadron tensor reads \cite{Collins:1989gx,Collins:2011zzd,Ji:2004wu} (here for SIDIS)
\begin{eqnarray}\label{TMD_factoriz}
W^{\mu\nu}(Q,q_{hT})=\sum_f H^{\mu\nu}(Q,\mu)\int \frac{d^2b_T}{(2\pi)^2}e^{-iq_{hT}b_T}F_{f/A}(x,b_T;\mu,\zeta_A)D_{B/\bar
f}(z,b_T;\mu,\zeta_B)+Y,
\end{eqnarray}
where $H$ is the hard coefficient function, $F$($D$) is TMD PDF (FF), $x$ and $z$ are longitudinal parts of parton momenta, $\mu$ and $\zeta$
are scales of the factorization. The $Y$-term accumulates corrections significant at $q_{hT}\sim Q$.

TMDs depend on four parameters. So, the dependence on factorization scales $\mu$ and $\zeta$ is given by renormalization group equation (RGE)
and Collins-Soper-Sterman (CSS) equation \cite{Collins:1984kg}. These dependencies have been intensively studied during last years (see
e.g.\cite{Aybat:2011zv,Aybat:2011ge,Bacchetta:2013pqa,Echevarria:2012pw}). While the dependence of TMDs on the parameters $x$ and $b_T$ cannot
be extracted within perturbative QCD due to nonperturbative nature of hadron states. In this paper we concentrate on the $x-$ and
$b_T-$dependence of TMDs leaving $\mu-$ and $\zeta$-dependence aside. For simplicity, we also set aside polarization effects and consider only
non-polarized TMDs.

The explicit expression for TMD PDF has the form of a nonlocal operator sandwiched between hadron states. The parton fields are separated by the
space-like distance $\xi=(0^+,\xi^-,b_T)$ and equipped by a construction of Wilson lines. A typical representative is the quark operator for TMD
PDF (see e.g.\cite{Collins:2011zzd,Ji:2004wu,GarciaEchevarria:2011rb,Cherednikov:2007tw})
\begin{eqnarray}\label{QCD:Op}
&&O_q(x,b_T;\mu,\zeta)=\\\nn &&Z_q(\mu)S^{-\frac{1}{2}}(b_T,\zeta)\int\frac{d\xi^-}{2\pi}e^{-ix p^+\xi^-}\bar
q_r\(\frac{\xi}{2}\)W^\dagger\(\frac{\xi}{2},-\infty; n\)\frac{\gamma^+}{2}W\(-\frac{\xi}{2},-\infty; n\)q_r\(-\frac{\xi}{2}\),
\end{eqnarray}
where $q_r$ are renormalized quark fields, $W(a,b;n)$ is Wilson line from point $a$ to point $b$ along direction $n$ ($n^2=0$). The factors
$Z_q$ and $S$ are field renormalization constant and soft factor, respectively. This factors are singular and responsible for the cancelation of
ultraviolet and rapidity divergences.

The factorized expression (\ref{TMD_factoriz}) is suitable for the phenomenological application. However, usually another representation for TMD
is used. Following \cite{Collins:2011zzd} we call this representation as TMD with maximum perturbative content. In this representation TMDs are
given by
\begin{eqnarray}\label{F_Cfeg}
F(x,b_T;\mu,\zeta)=C\(x,b_T;\mu,\zeta\)\otimes f(x,\mu) e^{-g(z,b_T,\zeta)},
\end{eqnarray}
where $\otimes$ is the Mellin convolution in $x$, $C$ is the coefficient function, $f$ is the integrated PDF and $e^{-g}$ is the
non-perturbative factor. This is the general ansatz for TMDs widely used in phenomenology, although the particular details of the representation
differ between approaches (compare e.g. \cite{Aybat:2011zv,Sun:2013hua,Echevarria:2014rua}, for the recent applications see e.g.
\cite{Echevarria:2014rua,Aidala:2014hva,Anselmino:2013lza} and reference within).

The coefficient function in (\ref{F_Cfeg}) is obtained from the leading terms of operator product expansion (OPE) for the TMD operator
(\ref{QCD:Op}). We emphasize the fact that used OPE holds only in the region of small $b_T$. Thus one should impose a cutoff over $b_T$. The
typical size of cutoff is $b_{max}^2=0.5-2$ GeV$^{-2}$. This boundary is motivated by a convergence radius of perturbative expansion for the
coefficient function $C$.

In the representation (\ref{F_Cfeg}) the non-perturbative factor plays the central role. It accumulates the most significant portion of
information on $b_T$. One can resolve its $\zeta$-dependance with the help of evolution equations and present the function $g$ in the form (see
e.g.\cite{Collins:2011zzd,Aybat:2011zv})
\begin{eqnarray}\label{g_usual}
g(x,b_T,\zeta)=g_f(x,b_T)+g_K(b_T)\ln\frac{\zeta}{\zeta_0},
\end{eqnarray}
where $g_f$ and $g_K$ cannot be obtained in the model-independent way and should be fitted from experiment. The prevalent parametrization for
the functions $g_{f,K}$ is the Gaussian ansatz, $g_{f,K}\sim b^2_T/4B_T^2$. This parametrization results to reasonable description of data. The
typical size of Gaussian is about $B_T^2=0.2-0.6$ GeV$^{-2}$.

In any parametrization the non-perturbative function $g$ reduces to zero at $b_T\to0$. In this limit TMDs match integrated parton distributions.
Therefore, the representation (\ref{F_Cfeg}) describes the TMDs at small $b_T$ within the perturbative QCD, while at larger $b_T$ it is replaced
by unknown function. In the following we discuss to which limits the perturbative content of representation (\ref{F_Cfeg}) can be used and
possible way to extend these limits.

\section{Intrinsic scales of small $b_T$ OPE}

In this section we discuss the properties of the non-perturbative factor and OPE. In the following, we keep in mind the Gaussian ansatz for the
non-perturbative factor $g=b_T^2/4B_T^2$. Moreover, we use the expression (\ref{F_Cfeg}) (with Gaussian non-pertrubative factor) as a kind of
the standard, that perfectly describes the data. We make such conjecture due to the lack of theoretical methods for analysis of TMDs at
intermediate $b_T$. The similar arguments which we will present can be applied for any other parameterizations with the same general conclusion.
The only privilege of the Gaussian ansatz is its simplicity and popularity.

Let us consider the OPE which leads to the expression (\ref{F_Cfeg}) closely. It reads
\begin{eqnarray}\label{intro:O_G_nO_n}
O(x,b_T)=\sum_{n=0}^\infty C^{(T)}_n(x,b_T)\otimes O^{(T)}_n(x),
\end{eqnarray}
where the operators $O^{(T)}_n$ are proportional to the $n$'th power of transverse derivative, $O^{(T)}_n\sim \bar q\partial_T^n q$ and the
coefficient functions $C^{(T)}_n$ are proportional to $b_T^n$. We omit the factorization scales $\mu$ and $\zeta$ for brevity. In the absence of
interaction the right-hand-side of (\ref{intro:O_G_nO_n}) represents the Taylor series of the operator $O(x,b_T)$ at $b_T=0$, that we indicate
by superscript $T$. The coefficient function in equation (\ref{F_Cfeg}) is $C_0^{(T)}$ in this notation.

In fact, the series (\ref{intro:O_G_nO_n}) is a double expansion, because every coefficient function $C_n^{(T)}$ is a perturbative series.
Therefore, the series (\ref{intro:O_G_nO_n}) has two main intrinsic scales $b_{max}$ and $B_T$. The scale $b_{max}$ is the universal scale of
convergence for the perturbative expansion for coefficient functions. It is naturally connected with $\Lambda^{-1}_{QCD}$. The origin of the
scale $B_T$ is non-pertrubative, $B_T$ parameterizes some intrinsic dynamics of hadron.

Taking the hadron matrix element of (\ref{intro:O_G_nO_n}) one obtains TMD in the form
\begin{eqnarray}\label{F_taylor}
F(x,b_T)=C^{(T)}_0(x,b_T)\otimes f(x)+\sum_{n=1}^\infty C^{(T)}_n(x,b_T)\otimes f_n(x),
\end{eqnarray}
where $f_n$ are integrated PDFs of higher twists. Comparing expressions (\ref{F_taylor}) and (\ref{F_Cfeg}) we conclude that $C_n\otimes f_n\sim
b_T^n/B_T^n$. In other words, the higher terms of OPE are of the same order at $b_T\sim B_T$. We stress that there are no perturbative methods
to estimate the radius of convergence for OPE (\ref{intro:O_G_nO_n}), and that our conclusion on behavior of higher terms is based only on the
phenomenological significance of the non-perturbative factor.

The scale $B_T$ is generally smaller then the scale $b_{max}$. It shows that the standard approach does not use the maximal possible
perturbative range of $b_T$, due to inefficiency of the power expansion. It suggests to use another basis which would saturate OPE within the
perturbative range by the first terms. In ref.\cite{Vladimirov:2014aja} such a modified approach to TMDs has been suggested. In the following we
present the main points of \cite{Vladimirov:2014aja}.

\section{Small $b_T$ OPE in Laguerre basis}

The main idea of \cite{Vladimirov:2014aja} is to rearrange small $b_T$ OPE in order to simulate the the non-perturbative behavior. Choosing
suitable basis for OPE one can obtain any preassigned form of $b_T$-distribution already at the leading order. The perturbative corrections
would tend to fit the expansion to the ``true'' expression within the radius of perturbative convergence. The control of the convergence is to
be obtained from the comparison with experiment. Therefore, the operator basis should be taken such that its leading terms describes the
significant part of data. We call such an approach as \textit{phenomenologically motivated OPE}. Technically it goes in parallel to the standard
approach of ref.\cite{Collins:2011zzd} and does not spoil any evolution or other properties of TMDs.

There are no special restrictions on the operator basis. It should be transversally local, orthogonal and complete. These are general demands
which guaranty the uniqueness and existence of the decomposition.  Additionally, one can impose symmetry or other constraints, which follow from
the auxiliary guidelines. Within these assumptions one can choose any basis.

Let us assume that the small $b_T$ range of TMDs is described by the Gaussian behavior. For the description of such a leading behavior the best
choice is the basis of Laguerre polynomials $L_n$. We have
\begin{eqnarray}\label{intro:O_G*L}
O(x,|b_T|)=\sum_{n=0}^\infty C^{(L)}_{n}(x,b_T;B_T)\otimes O^{(L)}_{n}(x;B_T),
\end{eqnarray}
where $O^{(L)}_n\sim L_{n}(B_T^2\partial^2)$. The coefficient functions of Laguerre expansion are Gaussians
$$
C^{(L)}_{n}(x,b_T;B_T)\sim \(\frac{b_T^{2}}{B_T^2}\)^ne^{-b_T^2/B_T^2}+\mathcal{O}(\alpha_s),
$$
which follow from the Gaussian form of the generating function for Laguerre polynomials. Additional argument in favor of Laguerre polynomial
basis is that Laguerre polynomials are the only classical orthogonal polynomials on the range $b_T\in(0,\infty)$.

The parameter $B_T$ in (\ref{intro:O_G*L}) is introduced for the dimensional reason. In general, OPE is independent on this parameter although
its convergence properties of OPE are dependent on it. In particular, the Laguerre based OPE (\ref{intro:O_G*L}) turns to the standard Taylor
based OPE (\ref{intro:O_G_nO_n}) in the limit $B_T\to\infty$. However, the truncated series which is used in practice, is $B_T$ dependent.

The $n=0$ term of OPE (\ref{intro:O_G*L}) is proportional to the integrated PDF operator. At the same time the higher terms of Laguerre based
OPE represent the mixture of operators of different twists including the leading one. However, this observation does not worsen the approach
since the contribution of different operators are of the same order. One can be guided only by experimental data, and tune the parameter $B_T$
such that $n>0$ terms give negligible contribution.

One may say that the change of the operator basis redistribute the power corrections between the terms of OPE. In such a picture the parameter
$B_T$ can be viewed as a handle which controls the amount of redistributed power corrections, while Laguerre polynomials modulate the
redistribution to the Gaussian shape.

In the free theory the suggested scheme does not add anything new to the standard description of TMDs with Gaussian non-perturbative factor. The
new results and predictions of the scheme appear with the loop-corrections to the coefficient function. The corrections produces the deviation
of the functional form of coefficient function from the free-theory limit. In the Taylor-like OPE the corrections can contain only the
logarithms of $b_T$. In the Laguerre based expansion, the other type of corrections are possible, e.g. power corrections and exponentials. These
corrections are of the special interest, because they show the perturbative deviation from the Gaussian ansatz. At the same time these
corrections are small within the perturbative range $b<b_{max}$ and do not spoil the general picture.

At large $b_T$ (i.e. $b_T>b_{max}$) the convergence of OPE is not controlled. Therefore, the usage of Laguerre (or any other) basis does not
eliminate the non-perturbative factor. However, one can expect that this new non-perturbative factor is much closer to unity within perturbative
range in comparison to the standard non-perturbative factor.

\section{Modified expression for TMD PDF}

\begin{figure}[t]
\includegraphics[width=0.3\textwidth]{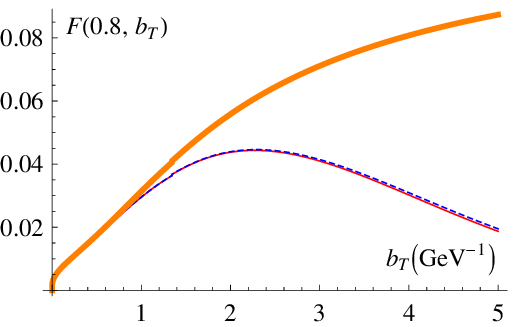}~~~\includegraphics[width=0.3\textwidth]{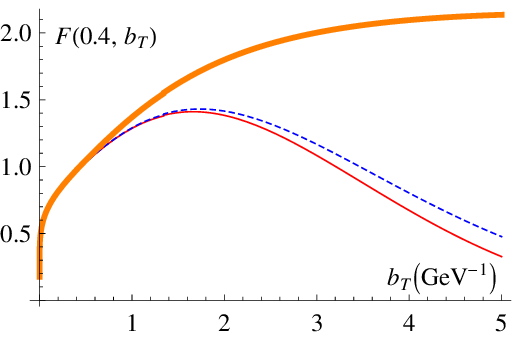}~~~\includegraphics[width=0.3\textwidth]{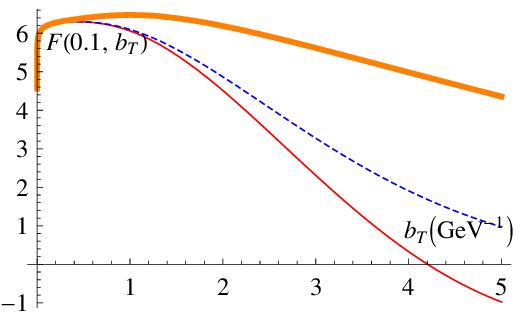}
\caption{Plots of the first terms Laguerre based expansion of TMD PDF (red curve) at different values of $x$ ($x=0.8,0.4,0.1$ from left to right
panels) at $b_{max}=1$GeV$^{-1}$ and $B_T^2=0.24$GeV$^{-2}$. The thick-orange curves are the first term of TMD PDF Taylor based expansion. The
blue-dashed curves are the first term of TMD PDF Taylor based expansion multiplied by the non-perturbative factor $\exp(-b_T^2/4B_T^2)$. The
evolution exponent is omitted.} \label{fig:TMDs}
\end{figure}

Taking the hadron matrix element of the Laguerre based OPE (\ref{intro:O_G*L}) we obtain the modified expression for the TMD PDF. It reads
\begin{eqnarray}\label{QCD:F_f}
F_{q/H}(x,b_T;\mu,\zeta)&=&\sum_{j}\int_x^1\frac{dz}{z}C^{(L)}_{q/j}\(\frac{x}{z},b_T;\mu,\zeta\)f_{j/H}(z,\mu)+\mathcal{O}_1,
\end{eqnarray}
where $f$ is the integrated PDF. The symbol $\mathcal{O}_1$ denotes the order of eliminated contribution. As we have discussed in the previous
section the estimation of $\mathcal{O}_1$ is impossible within the perturbative QCD. In the following we suppose that $\mathcal{O}_1$ is
negligible in comparison with the first term of (\ref{QCD:F_f}) within the perturbative range.

The coefficient functions $C^{(L)}$ have been calculated at NLO in \cite{Vladimirov:2014aja} and read
\begin{eqnarray}\label{QCD:Laguerre_Cqq_n_0}
C^{(L)}_{q/q}(x,b_T,\mu,\zeta)&=&e^{-\frac{b_T^2}{4B_T^2}}\delta(1-x)+
 \\ \nn &&
 2a_s C_Fe^{-\frac{x^2b_T^2}{4B_T^2}}\Bigg[-L_TP_{qq}(x)+\delta(\bar
x)\(\frac{3}{2}L_T-\frac{1}{2}L_T^2-\frac{\pi^2}{12}+L_T\ln\(\frac{\mu^2}{\zeta}\)\)+\bar x
\\\nn&&~~~~~~~~~~~~~~-\frac{\bar x x^2}{4} \frac{b_T^2}{B_T^2}L_T\(\frac{x^2}{4}\frac{b_T^2}{B_T^2}-3\)+\frac{x^4\bar x}{8}\(\frac{b_T^2}{B_T^2}\)^2
-x^2\bar x\frac{b_T^2}{B_T^2} \Bigg]+\mathcal{O}(a_s^2),
\\\label{QCD:Laguerre_Cqg_n_0}
C^{(L)}_{q/g}(x,b_T,\mu,\zeta)&=& 2a_s e^{-\frac{x^2b_T^2}{4B_T^2}}\(-P_{qg}(x)L_T+2x\bar
 x\)+\mathcal{O}(a_s^2),
\end{eqnarray}
where $a_s=g^2/(4\pi)^2$, $L_T=\ln\(b^2_T\mu^2/4e^{-2\gamma_E}\)$ and $P$ are the corresponded DGLAP kernels
$$
P_{qq}(x)=\(\frac{1+x^2}{1-x}\)_+,~~~~~~P_{qg}(x)=1-2x\bar x.
$$
At $B_T\to \infty$ these expressions reveal the standard expressions for the matching coefficients of TMD PDF to integrated PDF
(\cite{Collins:2011zzd,Aybat:2011zv,Echevarria:2012pw}).

In fig.1 we show the comparison of Taylor based expansion (\ref{F_taylor}) and Laguerre based expansion (\ref{QCD:Laguerre_Cqq_n_0}) (both
without non-perturbative factor) with the standard expression (\ref{F_Cfeg}). In contrast to the Taylor expansion, the Laguerre expansion
reproduces TMD PDF in the wider range of $b_T$ as it was expected. At smaller $x$ the resulting distribution is broader, i.e. the slope of
Gaussian is smaller. This is very natural result which shows that at smaller $x$ partons are allowed to be farer from the centrum of hadron.

\section{Conclusion}
We suggest the modification of the standard approach to TMDs. The modification consists in the consideration of the small $b_T$ OPE (which is
the central part of the standard approach) in the modified operator basis. So, instead of power expansion we suggest to use the Laguerre
polynomial expansion. Within such a scheme the first term of OPE describes the data in the wider range of $b_T$ in comparison to the power
expansion.

Such an approach is systematic, in the sense that it allows one to take into account quantum corrections systematically, and make comparison
with the standard approach at every step of the consideration. This approach does not violate the standard basic properties of TMD and TMD
factorization theorems, such as evolution equation, CSS-equation, convergence of the perturbative series and other. The modified expansion has
the same status as the standard one, since the size of corrections to both expressions cannot be estimated within perturbative QCD.

The choice of Laguerre polynomials as a basis for OPE is dictated by their simplicity and the Gaussian form of resulting coefficient function
(which is often used as phenomenological ansatz for TMDs). One can use another orthogonal and complete basis which would lead to different
behavior of coefficient function, with all the rest properties of TMDs survived. In the absence of theoretical constraints the choise of the
basis can be done only by comparison with the experimental data.

\acknowledgments The work is supported in parts by the European Community-Research Infrastructure Integrating Activity Study of Strongly
Interacting Matter" (HadronPhysics3, Grant Agreement No. 28 3286) and the Swedish Research Council grants 621-2011-5080 and 621-2010-3326.

\end{document}